\documentclass[12pt]{iopart}

\usepackage{iopams}  
\usepackage{setspace}
\expandafter\let\csname equation*\endcsname\relax
\expandafter\let\csname endequation*\endcsname\relax
\usepackage{amsmath}
\usepackage[margin=1.25in]{geometry}
\usepackage{graphicx}
\graphicspath{ {./figures/} }
\usepackage{subcaption}
\usepackage{lineno}
\usepackage[hidelinks]{hyperref}
\usepackage{hyperref}
\renewcommand{\abstractname}{\vspace{-\baselineskip}}
\usepackage{lipsum}
\usepackage{algorithm}
\usepackage{algpseudocode}
\usepackage{verbatim}  
\usepackage{bm}
\usepackage[english]{babel}
\usepackage{amsthm}
\theoremstyle{definition}
\newtheorem{definition}{Definition}[section]
\usepackage{float}
\usepackage{calc}  
\usepackage{enumitem}
\usepackage{varwidth}
\usepackage{tasks}
\setlist{  
  listparindent=\parindent,
  parsep=0pt,
}
\setlist{leftmargin=20mm}
\usepackage[style=nejm, 
citestyle=numeric-comp,
sorting=none]{biblatex}
\usepackage{amsfonts}
\begin{document}

\title[Clustering by Contour coreset and variational quantum eigensolver]{Clustering by Contour coreset and variational quantum eigensolver}

\author{Canaan Yung$^1$ \& Muhammad Usman$^{2,3}$}
\begin{indented}
\item[]$^1$School of Computing and Information Systems, The University of Melbourne, Parkville, VIC, 3010, Australia
\item[]$^2$School of Physics, The University of Melbourne, Parkville, VIC, 3010, Australia
\item[]$^3$Data61, CSIRO, Clayton Victoria Australia.
\end{indented}
\ead{canaany@student.unimelb.edu.au}

\begin{abstract}
Recent work has proposed solving the k-means clustering problem on quantum computers via the Quantum Approximate Optimization Algorithm (QAOA) and coreset techniques. Although the current method demonstrates the possibility of quantum k-means clustering, it does not ensure high accuracy and consistency across a wide range of datasets. The existing coreset techniques are designed for classical algorithms and there has been no quantum-tailored coreset technique which is designed to boost the accuracy of quantum algorithms. In this work, we propose solving the k-means clustering problem with the variational quantum eigensolver (VQE) and a customised coreset method, the Contour coreset, which has been formulated with specific focus on quantum algorithms. Extensive simulations with synthetic and real-life data demonstrated that our VQE+Contour Coreset approach outperforms existing QAOA+Coreset k-means clustering approaches with higher accuracy and lower standard deviation. Our work has shown that quantum tailored coreset techniques has the potential to significantly boost the performance of quantum algorithms when compared to using generic off-the-shelf coreset techniques.

\end{abstract}

%
%
%
%
%

\section{Introduction}
Quantum computing is an emerging paradigm for solving computationally intensive problems by exploiting the special properties of quantum mechanics such as superposition and entanglement in the algorithm design. As the current generation of quantum computers are noisy, they offer limited capabilities to handle practical problems. Furthermore, the access to only a limited number of qubits allows relatively small-sized problems to be solved on the available quantum devices. In order to address practical problems, one possible approach adopted in recent literature is pre-processing classical data to reduce its size, which can then be handled by a quantum algorithm. A salient example of such technique is the application of coreset for classical data reduction integrated with quantum approximate optimization algorithm (QAOA) which allowed quantum computers to address k-means clustering problem with moderate accuracies \cite{harrow2020small, tomesh2021coreset, qu2022performance}. Coreset is a data reduction technique approximating a dataset into a smaller subset with weights for solving optimization problems \cite{harrow2020small}. This hybrid quantum-classical approach converts the 2-means clustering objective function into a weighted MAX-CUT problem and solves it on a quantum computer using QAOA. It was found that the hybrid approach had a lower accuracy than the pure classical computing approach, with around 10\% difference in performance.

The current approach of using quantum computing for clustering involves combining pre-existing coreset techniques with QAOA \cite{harrow2020small, tomesh2021coreset}. However, this method has poor accuracy because the coreset methods are not tailored for quantum algorithms, and QAOA requires complex circuits for optimization. The coreset constructions designed for classical clustering problems are incompatible with quantum computing because they rely on probabilistic sampling and do not adequately sample datasets with limited coreset points \cite{bachem2018scalable, bachem2017practical, bachem2018one}. This lack of thorough sampling weakens data approximation. Furthermore, QAOA struggles with parameter optimization under NISQ because of the need for deeper circuits. As an alternative, variational quantum eigensolver (VQE) is a variational algorithm that has yet to be explored in the context of clustering. By creating a customized coreset method paired with VQE, we have shown that it is possible to achieve higher accuracies with consistency across a variety of data distributions in solving the k-means clustering problem.

To the best of our knowledge, there is no coreset designed explicitly for quantum computing. Therefore, we developed a new technique hereby labelled as Contour coreset for 2-means clustering on quantum computers. It approximates large classical datasets with limited points by evenly distributing coreset points across the whole dataset, which is helpful for quantum computing with limited qubits. Moreover, the Contour coreset construction process utilize deterministic sampling which does not involve randomness, ensuring accurate and reliable clustering performance when integrated with the quantum approach. When used alongside VQE, the Contour coreset outperforms existing coresets in solving the 2-means clustering problem by achieving higher accuracy of over 10\% and lower standard deviation of up to 0.1.

Furthermore, we have used a first-order Taylor Approximation to mathematically derive the Hamiltonian for the VQE circuit for handling unevenly distributed data in quantum devices. Compared to the zero-order Taylor Approximation used in previous research experiments, the first-order approximation improves clustering accuracy by around 5\% by reflecting the uneven data ratio in the Hamiltonian term \cite{tomesh2021coreset, qu2022performance}. We conducted simulations using Contour coreset and VQE with first-order Taylor approximated Hamiltonian on synthetic and real-world datasets. We searched for the optimal settings for the VQE circuit, such as the optimizer for the quantum circuit and the number of repetitions for rotation and entanglement blocks under the implementation of Qiskit. Compared to other off-the-shelf coresets, our results showed that the Contour coreset reduces construction time and improves accuracy by 10\% when combined with VQE for solving the 2-means clustering problem. In addition, we achieved higher accuracy by around 10\% and up to 0.1 standard deviation reduction compared to previous literature's methods involving QAOA and other coresets.

This paper is organized as follows: Section \ref{exp} describes the simulation setup and the parameters tested. Section \ref{contourcoreset} introduces the Contour coreset and compares its performance to existing coresets. Section \ref{result} presents the result analysis. Finally, we conclude with our research discussion and conclusion in sections \ref{discussion} and \ref{conclusion}.

\section{Simulation setup}\label{exp}

\subsection{Proposed framework}

\begin{figure}
    \centering
    \includegraphics[width=\textwidth]{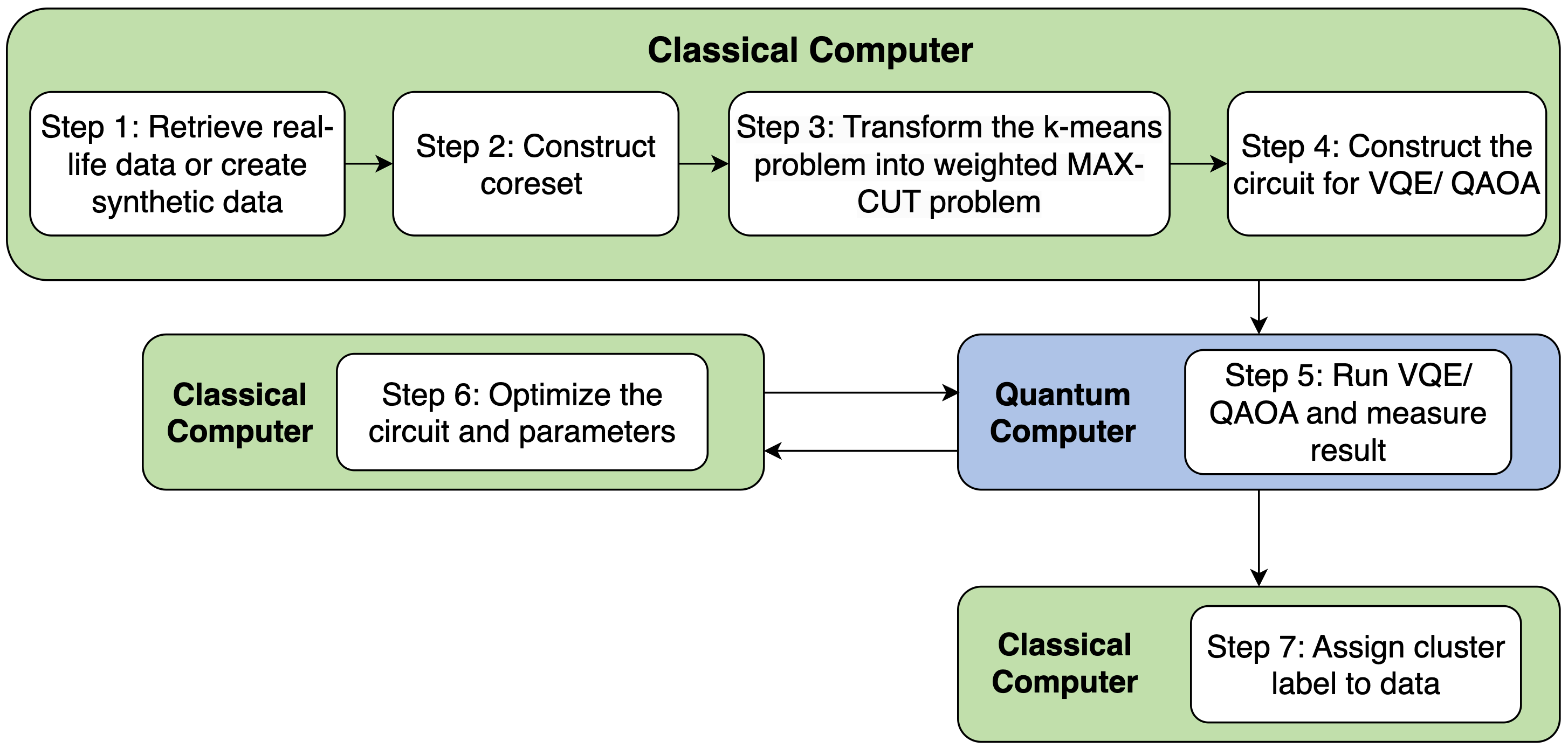}
    \caption[Framework flowchart]{Framework flowchart. The green and blue areas represent steps done on classical and quantum computers, respectively.}
    \label{experiment protocol}
\end{figure}

\begin{figure}
    \centering
    \includegraphics[width=\textwidth]{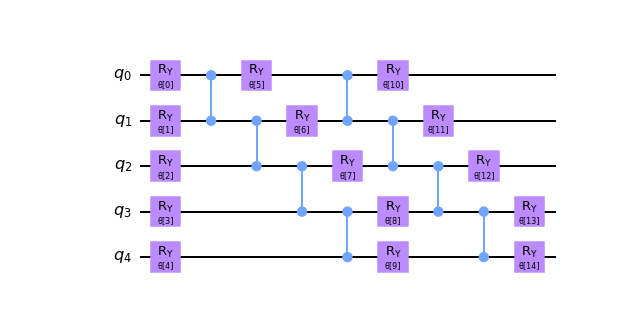}
    \caption[VQE circuit]{VQE circuit. This circuit consists of 5 qubits with 2 repetition of rotation and entanglement blocks.}
    \label{vqecircuit}
\end{figure}

\begin{figure}
    \centering
    \includegraphics[width=\textwidth]{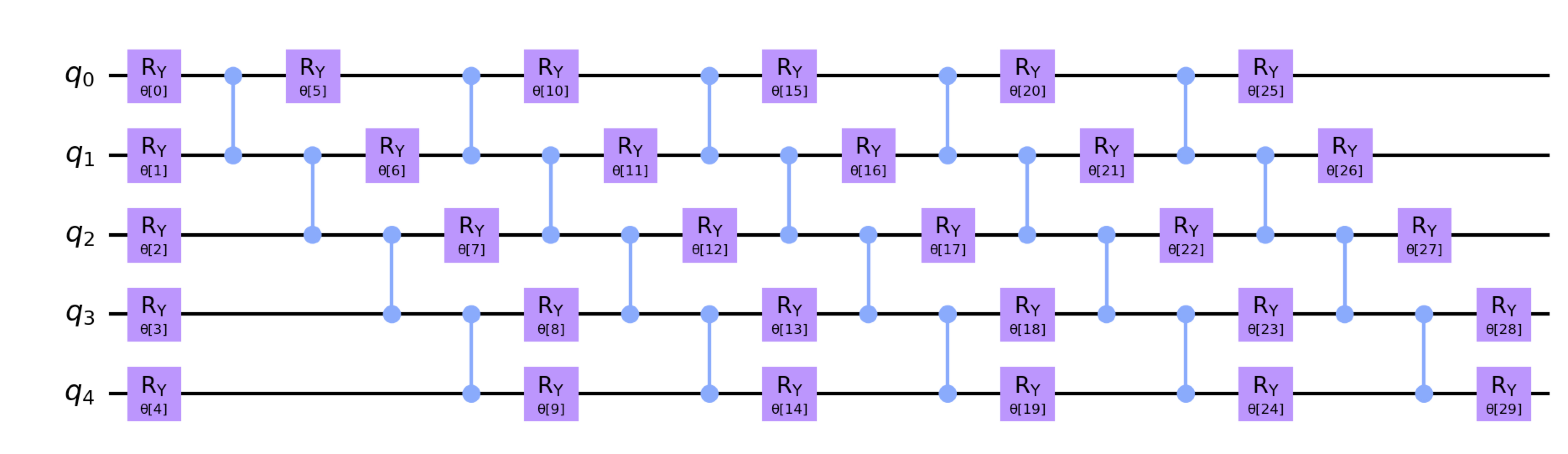}
    \caption[QAOA circuit]{QAOA circuit. This circuit consists of 5 qubits with 5 repetition of rotation and entanglement blocks.}
    \label{QAOAcircuit}
\end{figure}

Figure \ref{experiment protocol} illustrates the framework flowchart for solving the 2-means clustering problem with coresets and variational quantum algorithms. The simulation framework is explained as follows:

\begin{enumerate}

   \item[Step 1:] \textbf{Retrieve real-world data or create synthetic data.} We will test using real-world data and generate artificial data with Sklearn to test for extreme distribution imbalances.
   \item[Step 2:] \textbf{Coreset construction.} We analyze the performance of different types of coresets. The evaluation includes their construction time, clustering accuracy, and corresponding standard deviation. 
   \item[Step 3:] \textbf{Transformation of k-means clustering problem to weighted MAX-CUT problem.} The mathematical transformation is well established in past literature \cite{harrow2020small, tomesh2021coreset}. Our investigation will focus on the performance of creating the Hamiltonian through the zeroth and first-order Taylor approximations. The mathematical proof for Hamiltonian formation can be found in section \ref{maths proof for 2-means to max cut}. Note that the derivation of the second-order Taylor approximated Hamiltonian is new in our work and has not been performed in the past literature.
   \item[Step 4:] \textbf{VQE or QAOA circuit construction.} We will implement the VQE circuit using the Hamiltonian deduced in Step 3 (Figure \ref{vqecircuit}). We will also assess the performance of the QAOA circuit, utilizing the optimal parameters identified in previous research (Figure \ref{QAOAcircuit}).
   \item[Step 5:] \textbf{Run VQE or QAOA on quantum computers and result measurement.} We will be utilizing two quantum computer simulators, namely "ibm\_Vigo" and "qasm\_simulator". Additionally, we will examine how the circuit responds to different levels of depolarization noise.
   \item[Step 6:] \textbf{Circuit and parameters optimization.} The system parameters will be improved using local and global optimizers based on the measurement results. This process will be repeated with step 5 until the optimizer converges or reaches the maximum number of optimization steps.
   \item[Step 7:] \textbf{Assign the cluster label to all data.} We label data in the coresets using the measurement results of the VQE/QAOA circuit in step 6. Then, we calculate the mean of the coreset points in each class and find the two cluster centroids. Lastly, we assign each data point in the original classical dataset to the nearest centroid for clustering based on Euclidean distance.
   
\end{enumerate}

\subsection{Datasets}

\begin{figure}
\makebox[\linewidth][c]{%
\begin{subfigure}[b]{.6\textwidth}
\centering
\includegraphics[width=.95\textwidth]{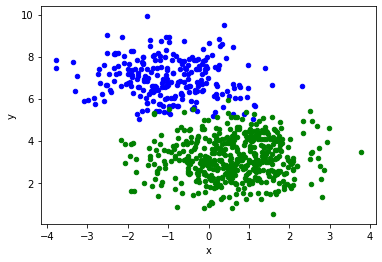}
\caption{Uneven 1:2}
\end{subfigure}%
\begin{subfigure}[b]{.6\textwidth}
\centering
\includegraphics[width=.95\textwidth]{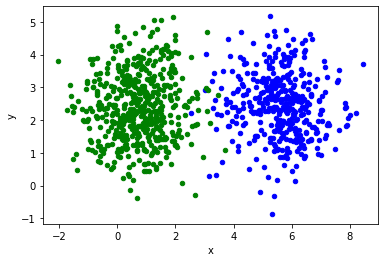}
\caption{Uneven 4:5}
\end{subfigure}%
}\\
\makebox[\linewidth][c]{%
\begin{subfigure}[b]{.6\textwidth}
\centering
\includegraphics[width=.95\textwidth]{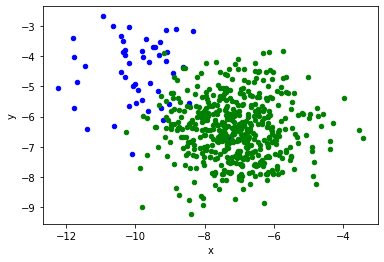}
\caption{Uneven 1:10}
\end{subfigure}%
}
\caption[Examples of Uneven distributed datasets]{Graphical illustration of isotropic Gaussian blobs with uneven distribution. The blue and green dots represent data from two clusters. Subfigure (a), (b) and (c) show the datasets with data ratios 1:2, 4:5 and 1:10 between the two clusters respectively.}
\label{unevendata}
\end{figure}

We tested our quantum approach on two types of datasets - synthetic and real-world. The synthetic datasets were produced using Sklearn library, which generates isotropic Gaussian blobs with uneven distribution. We created a range of datasets with two clusters, two cluster centers with a standard deviation of one, and different data ratios: 1:2, 4:5, and 1:10, which we named "Uneven 1:2", "Uneven 4:5", and "Uneven 1:10" (Figure \ref{unevendata}). It is worth noting that each simulation run used a new random seed to form the dataset. Meanwhile, we obtained numerous datasets from the UC Irvine Machine Learning repository for real-world datasets \cite{ucirvine}.

\subsection{Quantum circuit parameters optimization and coresets comparing}

This section will cover the quantum algorithm parameters and coresets we are testing to achieve optimal clustering performance with the proposed framework.

Our main objective is to evaluate and compare the effectiveness of two quantum algorithms: QAOA and VQE. We thoroughly analyzed previous research and developed a QAOA circuit to obtain optimal clustering outcomes. While VQE is typically used for quantum chemistry, there is no research on its performance in addressing the 2-means clustering problem with coresets \cite{li2019variational, hempel2018quantum}. Therefore, we will focus on identifying the best configurations for the VQE circuit that can produce an accurate clustering results. Lastly, we will investigate whether VQE can outperform QAOA when solving the 2-means clustering problem with coreset.

Additionally, we will test various parameter configurations for the quantum circuit. One primary focus is on the zeroth and first-order Taylor Approximation in generating the target Hamiltonian. Equations for the Hamiltonian were previously derived using the zeroth and first-order Taylor Approximation, but their performance in VQE and unevenly distributed datasets was not tested. \cite{tomesh2021coreset}. This leaves uncertainty about to what extent the clustering accuracy will be impacted using different orders of Taylor approximated Hamiltonians. In addition, we will optimize the quantum circuit by tuning the optimizer used, the number of repetitions, the number of qubits, and the entanglement strategies.

Finally, we will examine the effectiveness of four coresets: BFL16, ONESHOT, Lightweight, and our proposed Contour coreset. We compared the Contour coreset with these three coresets since they are developed for k-means clustering and utilized in investigating the quantum approach with QAOA in the past literature \cite{tomesh2021coreset, qu2022performance}.

\subsection{Analysis method}

The "Number of points correctly assigned" equals the number of points assigned to the same cluster by the classical and hybrid methods. The definition is as follows:

$$Accuracy = \frac{\text{Number of points correctly assigned}}{\text{Total number of points in dataset}}$$

The "Number of points correctly assigned" equals the number of points assigned to the same cluster by both methods. It is crucial to note that our implementation of assessing the effectiveness of coreset and VQE in solving k-means clustering differs from the previous studies mentioned in \cite{qu2022performance}. 

\begin{algorithm}
\caption{Accuracy calculation from past literature}\label{alg:Accuracy}
\begin{algorithmic}[1]
\Require $n \gets \text{no. of maximum iterations}, \text{threshold} \gets \text{minimum accuracy}$
\State $i = 0$
\While{$i \leq n$ or \text{highest accuracy} $<$ threshold}
\State $accuracy \gets \text{Clustering by VQE and coreset with desired variables}$
\If{\text{highest accuracy is null}}
    \State $\text{highest accuracy} \gets accuracy$
\EndIf
\If{$accuracy > \text{highest accuracy}$}
    \State $\text{highest accuracy} = accuracy$
\EndIf
\EndWhile
\State \Return \text{highest accuracy}
\end{algorithmic}
\end{algorithm}

Algorithm \ref{alg:Accuracy} illustrates the clustering accuracy calculation in past literature \cite{qu2022performance}. However, this method may result in overestimated accuracy measurements. The issue arises because the experimental data is only recorded when the clustering accuracy exceeds a certain threshold (line 2 of Algorithm \ref{alg:Accuracy}). Consequently, the program will remain in a while loop (lines 2-10) until it reaches the maximum iterations if the threshold is not met. 

The trial and error process for recording the clustering accuracy in Algorithm \ref{alg:Accuracy} can lead to an unintentional increase in the recorded clustering accuracy. This is because quantum errors may impact the measurement outcomes differently in each run, resulting in different clustering accuracy values. Moreover, since the coreset sampled in the same data may not be identical due to randomness in its construction, the Hamiltonian approximation and coreset formed in each run are different. This further contributes to the variation in the clustering accuracy obtained. Combined with the threshold in Algorithm \ref{alg:Accuracy} line 2, all these factors can lead to overestimating the accuracy. In real-life scenarios, it is not possible to cross-check the clustering result obtained by the quantum algorithm to the actual data labels (Algorithm \ref{alg:Accuracy} line 3) before finalizing the experimental result. The quality of the recorded accuracy should not be a concern during the intake of experimental data.

Our approach for accurate implementation is based on qubits' states from our quantum variational algorithm. Multiple independent simulations will be conducted to calculate average accuracy and reduce randomness. We will also examine standard deviation to ensure stability for practical use.

We will evaluate the Contour coreset by analyzing its construction time and plotting the position of coreset points. It is essential to minimize the construction time for efficiency and scalability. Furthermore, visualizing the Contour coreset points graphically will aid in comprehending their data approximation and evaluating if they sample data points across the two cluster groups.

\section{Contour Coreset -- A new technique for quantum method}\label{contourcoreset}

We will begin by explaining the incompatibility of the existing coreset to quantum computing, which is the motivation behind developing the Contour coreset. Then we will outline the construction algorithm of the Contour coreset and compare its performance with other coresets in solving the clustering problem.

\subsection{Incompatibility of existing coreset to the quantum approach}

We will focus on two coresets, BFL16 and ONESHOT, which have been used in the past literature to solve the k-means clustering problem with QAOA \cite{qu2022performance, tomesh2021coreset}. BFL16 and ONESHOT are well-established coreset methods used in classical literature for k-means clustering \cite{bachem2017practical, bachem2018scalable}. They are bounded by the ($\epsilon,k$)-coreset criterion for stable and optimal performance. Thus, the number of required coreset points increases with the dataset size for achieving optimal data sampling. Contrarily, quantum computers of the current generation offer access to limited number of qubits, which means only a few coreset points can be utilized for the estimation. As a result, the performance of coresets can vary significantly, with standard deviations reaching over 0.1. While using more advanced quantum computers with more available qubits may seem like a solution, increasing the number of qubits can exacerbate incoherent errors and affect the quantum state measurement.

\begin{figure}
    \centering
    \includegraphics[width=0.7\textwidth]{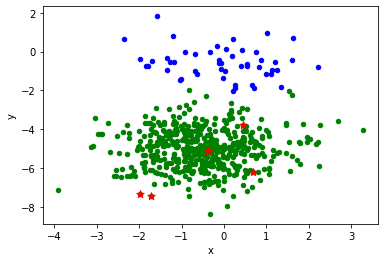}
    \caption[Imbalance sampling of Lightweight coreset]{Graphical illustration of imbalance Lightweight coreset sampling. The dataset consists of two clusters: blue and green dots, with an imbalanced data ratio of 1:10. The five Lightweight coreset points are plotted as red stars. These five coreset points are all clustered within the green majority group, and none are assigned to the blue minority cluster.}
    \label{incorrectLightweight}
\end{figure}

Additionally, coresets that use importance sampling can create imbalanced selection of the coreset. For example, the Lightweight coreset utilizes importance sampling in its construction algorithm. When the coreset size is drastically reduced, like in the quantum approach where only five points will be used, the data approximation may be imbalanced and not account for the minority cluster. Moreover, since Lightweight coreset sampling utilizes probabilistic sampling without replacement, a data point may be selected multiple times as a coreset point, resulting in fewer coreset points for data representation. The suboptimal distribution of coreset points can cause uncertainty in accuracy due to decreasing clustering performance and increasing standard deviation. Figure \ref{incorrectLightweight} illustrates an imbalance sampling of the Lightweight coreset on an "Uneven 1:10" dataset. All coreset points are plotted within the majority group, without any representation from the minority group. This is due to a lack of adequate coreset points, which fails to represent the overall probability distribution for data sampling accurately. 

\subsection{Contour coreset}

Due to the aforementioned limitations of standard coreset techniques, we developed a new coreset method which is labelled as Contour coreset. The Contour coreset algorithm is optimized for quantum algorithms and overcomes the limitations of current coreset algorithms. It performs well with limited coreset points, which is essential for quantum computation with limited qubits. It also provides balanced data approximation for unevenly distributed classical datasets. Our goal is to improve accuracy in clustering and reduce standard deviation when solving the 2-means clustering problem using variational quantum algorithms.

\begin{figure}
\makebox[\linewidth][c]{%
\begin{subfigure}[b]{.6\textwidth}
\centering
\includegraphics[width=.95\textwidth]{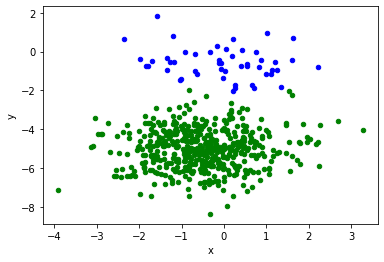}
\caption{Original dataset}
\end{subfigure}%
\begin{subfigure}[b]{.6\textwidth}
\centering
\includegraphics[width=.95\textwidth]{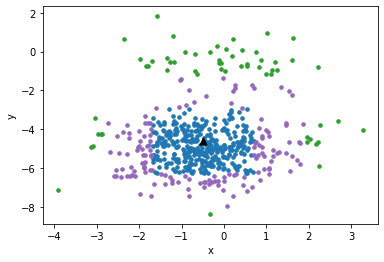}
\caption{Data division}
\label{datadivison}
\end{subfigure}%
}\\
\makebox[\linewidth][c]{%
\begin{subfigure}[b]{.6\textwidth}
\centering
\includegraphics[width=.95\textwidth]{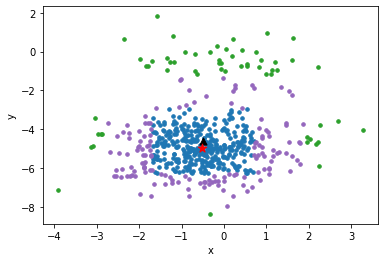}
\caption{First Contour coreset point}
\label{firststep}
\end{subfigure}%
\begin{subfigure}[b]{.6\textwidth}
\centering
\includegraphics[width=.95\textwidth]{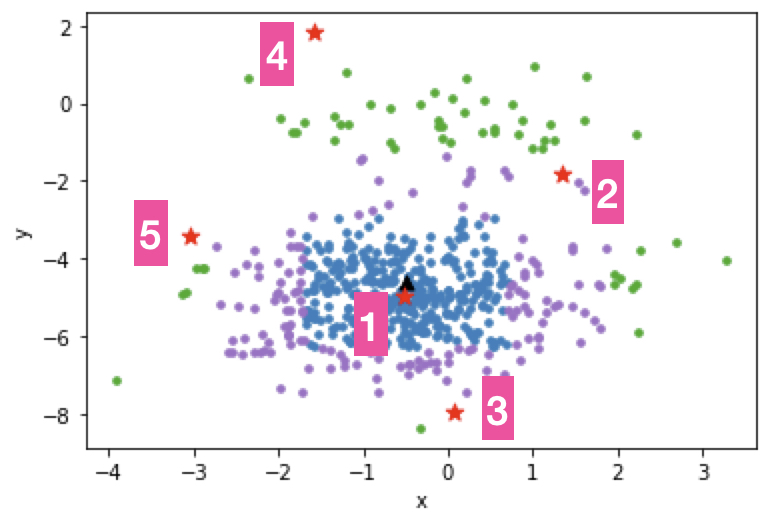}
\caption{Remaining Contour coreset points}
\label{second step}
\end{subfigure}%
}
\caption[Contour coreset]{Contour coreset: The subfigures show the procedure of constructing the Contour coreset for an imbalanced dataset. (a) displays the original dataset, consisting of two clusters, represented by blue and green data points. (b) shows the data division into three regions, with blue, purple, and green representing the innermost to outermost regions, respectively. The black triangle in the middle is the center of data. (c) illustrates the plotting of the first Contour coreset point through the varied Lightweight coreset approach, represented by a red star. (d) illustrates the remaining Contour coreset points (red stars) and their plotting order, indicated by numbers. We start by using the first point of the coreset as our reference point (i.e. first point in k-means++). Then, we apply k-means++ to plot the second coreset point in the purple area. Next, we use the first and second coreset points and apply k-means++ to plot the third point. Finally, for the fourth and fifth coreset points in the green area, we use the third coreset point as our reference and apply k-means++ to plot them both.}
\label{plotCC}
\end{figure}

The overview of the Contour coreset is as follows: The data is divided into regions (three regions by default) based on their distance from the data center. Then it uses k-means++ and Lightweight coreset algorithm techniques to deterministically plot evenly distributed coreset points in each region, preventing clustering in adjacent areas. The construction process is explained in detail in the following paragraphs, with Figure \ref{plotCC} providing a step-by-step illustration and pseudo-code is presented in Algorithm \ref{alg:contour}. 

\begin{algorithm}
\caption{Contour coreset construction}\label{alg:contour}
\begin{algorithmic}
\Require $\text{Set of data points $X$}, \text{Number of regions $k$} , \text{coreset size $m$}$
\State $\text{coreset} \gets [\;]$
\State $\text{coresetWeight} \gets [\;]$
\State $\text{dataInRegions} \gets \text{sortDataInRegions(X, k)}$ 
\State $\text{coresetPerRegions} \gets \text{coresetAssignment(dataInRegions, k, m)}$

\While{$\text{len(coreset)==0}$} 

\For{data in dataInRegions}

\If{data is elgibible}
\State $\text{coreset.append(firstCoreset)}$ 
\State coresetWeight.append(lightweightWeight(firstCoreset))
\State lastCoreset $\gets$ firstCoreset
\State \textbf{Break}
\EndIf
\EndFor
\EndWhile

\For{data in dataInRegions} 
\If{coresetPerRegions[data] $>$ 0} 
\State tmpCoreset, tmpWeights $\gets$ kMeansPlusPlus(data, coresetPerRegions[data], lastCoreset)
\State coreset.append(tmpCoreset)
\State coresetWeight.append(tmpWeights)
\EndIf
\EndFor
\State \Return \text{coreset, coreset weight}
\end{algorithmic}
\end{algorithm}

First, we sort the data into distinct regions (refer to Figure \ref{datadivison}). The division of data into regions is done based on their distance from the centre, and the number of regions can be customized. Setting the number of regions as an adjustable parameter gives us control over the approximate number of coreset points representing each region. The parameter ensures efficient data sampling for varying data sizes, data structures, and number of coreset points. For this study, we have chosen to create three regions since we are utilizing five qubits quantum computers, and we aim to assign approximately two coreset points per region for efficient data sampling. The data division process begins with determining the center of the original dataset, which is the average data point in each dimension. This is done by adding up the numerical values of all the data in each dimension and dividing it by the total number of data points. Next, we calculate the radius of each region. To do this, we first find the range of each dimension, which is the difference between the maximum and minimum values of data in that dimension. We then divide the range by the number of regions to determine the distance each region covers. Since we are sorting the data based on their distances from the center, we divide the distance of each region by half. By adding and subtracting the radii of the center, we can determine the range of each region. To define the regions, we use a formula that calculates the range of coverage for each dimension. The coverage of a region is determined by the center point plus or minus the region’s radii multiplied by the order of the region.

The second step involves determining the number of points assigned to each region. We assess region eligibility for assigning coreset points based on a minimum data threshold. The threshold equals to 

$$\text{threshold} = \Bigl\lfloor \frac{\text{total number of data points}}{\text{(number of regions)}*\text{(total number of coreset points)}}\Bigr\rfloor$$

The threshold determines the number of data points for each coreset point based on region and plot count. Now we can determine how many coreset points shall be assigned to each eligible region. The idea is that data points farther away from the mean have more influence and impact on clustering accuracy, which is also observed in previous literature \cite{bachem2018scalable}. We assign points iteratively, starting with the outermost area and moving towards the innermost area, each time assigning one point to each region, before repeating the process by starting over at the outermost area.

The unique method of assigning coreset points distinguishes the Contour coreset from trivial coresets, such as BFL16, ONESHOT and Lightweight coreset \cite{bachem2017practical, bachem2018scalable, bachem2018one}. The Contour coreset only assigns points to eligible regions, thus avoiding outlier data points or areas without data. This approach also prevents coreset points clustering in specific areas, which is a common problem with the Lightweight coreset. Additionally, the algorithm can improve clustering accuracy by assigning coreset points from the outermost to the innermost region in an iterative process while still providing a good approximation of all the data. Therefore, the Contour coreset algorithm produces a precise representation of the original dataset.

Now we can combine k-means++ and a variant of the Lightweight coreset approach in plotting the Contour coreset points \cite{arthur2007k, bachem2018one}. The first point is placed in the eligible region closest to the center (Figure \ref{firststep}). We use a varied Lightweight coreset approach to plot the data point with the shortest squared distance to the mean as our first coreset point. We then move on to the nearest eligible region (or stay in the same region if multiple coreset points are assigned) with unassigned points and plot the other coreset points using k-means ++ (Figure \ref{second step}). This process starts from inner to outer regions, with each new region starting from the last plotted coreset point as the k-means ++ starting point.

This approach offers two significant benefits. Firstly, it avoids randomness by consistently creating the same coreset for a dataset, thus improving clustering accuracy and reducing standard deviation. Many coreset algorithms in past literature involve randomness, such as importance sampling, which can affect the stability of the clustering performance \cite{bachem2017practical, bachem2018one, bachem2018scalable, lucic2016strong}. When using the quantum approach for clustering, errors such as approximations and noise from forming Hamiltonian and quantum operations can also impact achieving the most optimal results. All of these errors will be amplified and significantly impact the accuracy of the clustering result. Secondly, k-means ++ ensures that coreset points are not clustered nearby. Moreover, by taking into account the previous coreset point in the last region, the coreset points will not be adjacent to one another across regions.

Finally, we assign weights to coreset points using the Lightweight coreset algorithm. The distance from the mean determines each weight, calculated using the formula $\frac{1}{m \cdot q(\mathbf{x})}$, based on the Lightweight coreset \cite{bachem2018scalable}. The formula prioritizes clustering accuracy by giving more weight to points far from the mean.

\subsection{Comparison with other coresets}

\begin{figure}[]
    \centering
    \makebox[0pt]{\includegraphics[width=1.3\textwidth]{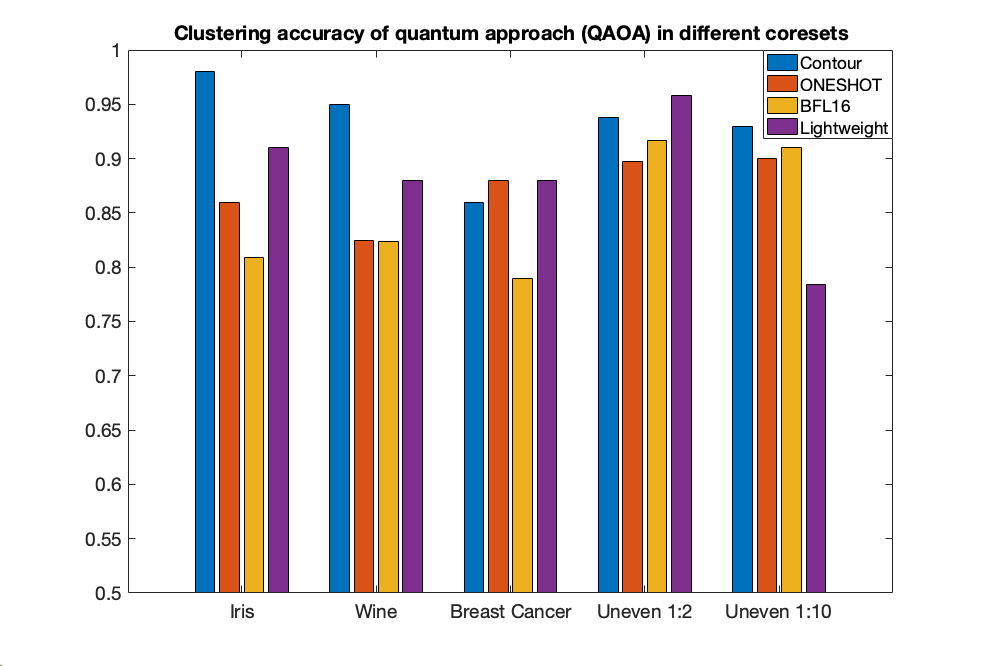}}
    \caption[Comparison of Contour coreset and other existing coresets in quantum approach (QAOA).]{Comparison of Contour coreset and other existing coresets in quantum approach (QAOA).}
    \label{qaoaContour}
\end{figure}

\begin{table}[h]
\centering
\begin{tabular}{|cccccc|}
\hline
\multicolumn{6}{|c|}{Standard deviation of the clustering accuracy acheived by QAOA and coresets} \\ \hline
\multicolumn{1}{|c|}{} &
  \multicolumn{1}{c|}{Iris} &
  \multicolumn{1}{c|}{Wine} &
  \multicolumn{1}{c|}{Breast Cancer} &
  \multicolumn{1}{c|}{Uneven 1:2} &
  Uneven 1:10 \\ \hline
\multicolumn{1}{|c|}{Contour} &
  \multicolumn{1}{c|}{0.16328} &
  \multicolumn{1}{c|}{0.102} &
  \multicolumn{1}{c|}{0.111} &
  \multicolumn{1}{c|}{0.109} &
  0.158 \\ \hline
\multicolumn{1}{|c|}{ONESHOT} &
  \multicolumn{1}{c|}{0.099} &
  \multicolumn{1}{c|}{0.134} &
  \multicolumn{1}{c|}{0.0524} &
  \multicolumn{1}{c|}{0.113} &
  0.162 \\ \hline
\multicolumn{1}{|c|}{BFL16} &
  \multicolumn{1}{c|}{0.1734} &
  \multicolumn{1}{c|}{0.122} &
  \multicolumn{1}{c|}{0.0914} &
  \multicolumn{1}{c|}{0.15} &
  0.154 \\ \hline
\multicolumn{1}{|c|}{Lightweight} &
  \multicolumn{1}{c|}{0.1238} &
  \multicolumn{1}{c|}{0.133} &
  \multicolumn{1}{c|}{0.0966} &
  \multicolumn{1}{c|}{0.164} &
  0.14 \\ \hline
\end{tabular}
\caption{Standard deviation of the clustering accuracy acheived by QAOA and coresets}
\label{SDQAOAcoresets}
\end{table}

We compare the Contour coreset's performance with other coresets for clustering problems. Our findings show that the Contour coreset has obtained a higher clustering accuracy and significantly reduced the standard deviation and construction time. The results are presented in Figure \ref{qaoaContour} and Table \ref{SDQAOAcoresets}. In most datasets, the Contour coreset has higher clustering accuracy than BFL16, ONESHOT, and Lightweight coreset. In particular, the Contour coreset stands out in the Iris and Wine datasets by up to 10\%. Besides, the Contour coreset has a higher minimum accuracy than other coresets, indicating better worst-case performance. It also has a smaller accuracy range and lower standard deviation than other coresets, hence a more consistent performance.

\begin{figure}
\makebox[\linewidth][c]{%
\begin{subfigure}[b]{.6\textwidth}
\centering
\includegraphics[width=.95\textwidth]{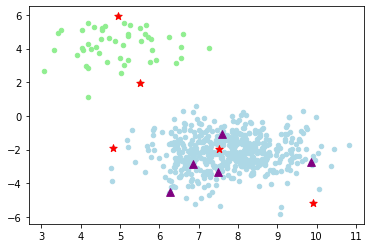}
\caption{Contour vs Lightweight}
\end{subfigure}%
\begin{subfigure}[b]{.6\textwidth}
\centering
\includegraphics[width=.95\textwidth]{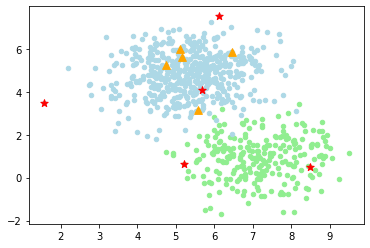}
\caption{Contour vs ONESHOT}
\end{subfigure}%
}
\caption[Coreset points comparison]{Coreset comparison: Two clusters in the dataset, one light blue and one green (a) Contour coreset points are red stars, Lightweight coreset are purple triangles. (b) Contour coreset points are red stars, ONESHOT coreset are orange triangles.}
\label{lightvsContour}
\end{figure}

We also analyzed 2D datasets to show the effectiveness of the Contour coreset in Figure \ref{lightvsContour}. We observe that the Lightweight coreset only clustered the majority light blue cluster and disregarded the minority green cluster. On the contrary, the Contour coreset accommodated both clusters. We noticed the same phenomenon in the ONESHOT coreset as it did not plot any coreset in the minority green cluster. Moreover, some of the ONESHOT coreset points were too close to each other, which hindered data approximation even more. These results indicate that the Contour coreset is capable of handling datasets with uneven distribution.

\begin{table}[]
\centering
\begin{tabular}{|ccccccc|}
\hline
\multicolumn{7}{|c|}{Table of number of instances and attributes of datasets} \\ \hline
\multicolumn{1}{|c|}{} &
  \multicolumn{1}{c|}{Iris} &
  \multicolumn{1}{c|}{Wine} &
  \multicolumn{1}{c|}{Breast Cancer} &
  \multicolumn{1}{c|}{Uneven 1:2} &
  \multicolumn{1}{c|}{Uneven 1:10} &
  Epilspsy \\ \hline
\multicolumn{1}{|c|}{Instances} &
  \multicolumn{1}{c|}{150} &
  \multicolumn{1}{c|}{178} &
  \multicolumn{1}{c|}{569} &
  \multicolumn{1}{c|}{750} &
  \multicolumn{1}{c|}{550} &
  11500 \\ \hline
\multicolumn{1}{|c|}{Attributes} &
  \multicolumn{1}{c|}{4} &
  \multicolumn{1}{c|}{13} &
  \multicolumn{1}{c|}{30} &
  \multicolumn{1}{c|}{2} &
  \multicolumn{1}{c|}{2} &
  179 \\ \hline
\end{tabular}
\caption[Table of number of instances and attributes of datasets]{Table of number of instances and attributes of datasets}
\label{datasetdim}
\end{table}

\begin{table}[]
\centering
\begin{tabular}{|ccccccc|}
\hline
\multicolumn{7}{|c|}{Time to construct coreset on a standard desktop computer (in seconds)} \\ \hline
\multicolumn{1}{|c|}{} &
  \multicolumn{1}{c|}{Iris} &
  \multicolumn{1}{c|}{Wine} &
  \multicolumn{1}{c|}{Breast Cancer} &
  \multicolumn{1}{c|}{Uneven 1:2} &
  \multicolumn{1}{c|}{Uneven 1:10} &
  Epilspsy \\ \hline
\multicolumn{1}{|c|}{Contour} &
  \multicolumn{1}{c|}{0.000210} &
  \multicolumn{1}{c|}{0.000212} &
  \multicolumn{1}{c|}{0.000284} &
  \multicolumn{1}{c|}{0.000284} &
  \multicolumn{1}{c|}{0.000293} &
  0.004423 \\ \hline
\multicolumn{1}{|c|}{ONESHOT} &
  \multicolumn{1}{c|}{0.150} &
  \multicolumn{1}{c|}{0.183} &
  \multicolumn{1}{c|}{0.716} &
  \multicolumn{1}{c|}{0.957} &
  \multicolumn{1}{c|}{0.687} &
  26.0 \\ \hline
\multicolumn{1}{|c|}{BFL16} &
  \multicolumn{1}{c|}{0.0444} &
  \multicolumn{1}{c|}{0.0519} &
  \multicolumn{1}{c|}{0.150} &
  \multicolumn{1}{c|}{0.197} &
  \multicolumn{1}{c|}{0.144} &
  3.04 \\ \hline
\multicolumn{1}{|c|}{Lightweight} &
  \multicolumn{1}{c|}{0.000290} &
  \multicolumn{1}{c|}{0.000293} &
  \multicolumn{1}{c|}{0.000293} &
  \multicolumn{1}{c|}{0.000293} &
  \multicolumn{1}{c|}{0.000293} &
  0.000293 \\ \hline
\end{tabular}
\caption[Table of coresets' construction time]{Table of coresets' construction time.}
\label{timeCoreset}
\end{table}

Lastly, we measured the time it takes to construct the coreset by executing the construction algorithms on a MacBook Pro 2021, equipped with Apple M1 Max and 64 GB memory. We measured the construction time by calculating the time elapsed from running the coreset construction algorithm to acquiring the coreset position vectors and their respective weights. The number of instances and attributes of the simulation datasets are displayed in Table \ref{datasetdim}, and the construction time of coresets for each dataset is shown in Table \ref{timeCoreset}. In most real-world datasets, the Contour coreset is highly effective with a construction time of only 0.0002 seconds compared to the 0.1 to 1 second required by ONESHOT and BFL16. In unevenly distributed synthetic datasets, the BFL16 and ONESHOT coresets take significantly longer to construct, while the Contour coreset is unaffected. Even in the largest dataset, Epilepsy, the Contour coreset's construction time only increased slightly to 0.004 seconds, which is negligible. Therefore, the Contour coreset is a better option for efficient coreset construction than other existing options.

\section{Results and Discussions}\label{result}

The optimal settings for our proposed quantum approach in solving the 2-means clustering problem are presented in Table \ref{optSettings}. Note that we will only manipulate the independent variable we are testing in these simulations while holding all other simulation settings constant as outlined in Table \ref{optSettings}. We conducted tests on the parameters of the quantum circuit using Qiskit's available adjustable parameters. Also, we tested all of Qiskit's local optimizers and global optimizers developed by NLopt. The coreset options we tested include the Contour coreset, BFL 16, ONESHOT, and Lightweight.

\begin{table}[H]
\centering
\begin{tabular}{|ll|}
\hline
\multicolumn{2}{|c|}{\begin{tabular}[c]{@{}c@{}}Optimal settings for our proposed quantum approach \end{tabular}} \\ \hline
\multicolumn{1}{|l|}{Coreset}                                                   & Contour coreset                  \\ \hline
\multicolumn{1}{|l|}{Size of coreset/ Number of qubits}                         & 5                                \\ \hline
\multicolumn{1}{|l|}{Quantum Variational Algorithm}                             & Variational Quantum Eigensolver  \\ \hline
\multicolumn{1}{|l|}{Hamiltonian Formation}                                     & First Order Taylor Approximation \\ \hline
\multicolumn{1}{|l|}{Optimizer for quantum circuit}  & \vtop{\hbox{\strut Improved Stochastic Ranking}\hbox{\strut Evolution Strategy optimizer (ISRES)}}\\ \hline
\multicolumn{1}{|l|}{Entanglement}                                              & Linear                \\ \hline
\multicolumn{1}{|l|}{\vtop{\hbox{\strut Number of repetition for}\hbox{\strut rotation and entanglement blocks}}} & 2                                \\ \hline
\end{tabular}
\caption[Optimal settings for our proposed quantum approach with coreset and quantum variational algorithm]{Optimal settings for our proposed quantum approach with coreset and quantum variational algorithm}
\label{optSettings}
\end{table}

\subsection{Comparison of Coreset Techniques with VQE Algorithm
}

\begin{figure}[]
    \centering
    \makebox[0pt]{\includegraphics[width=1.2\textwidth]{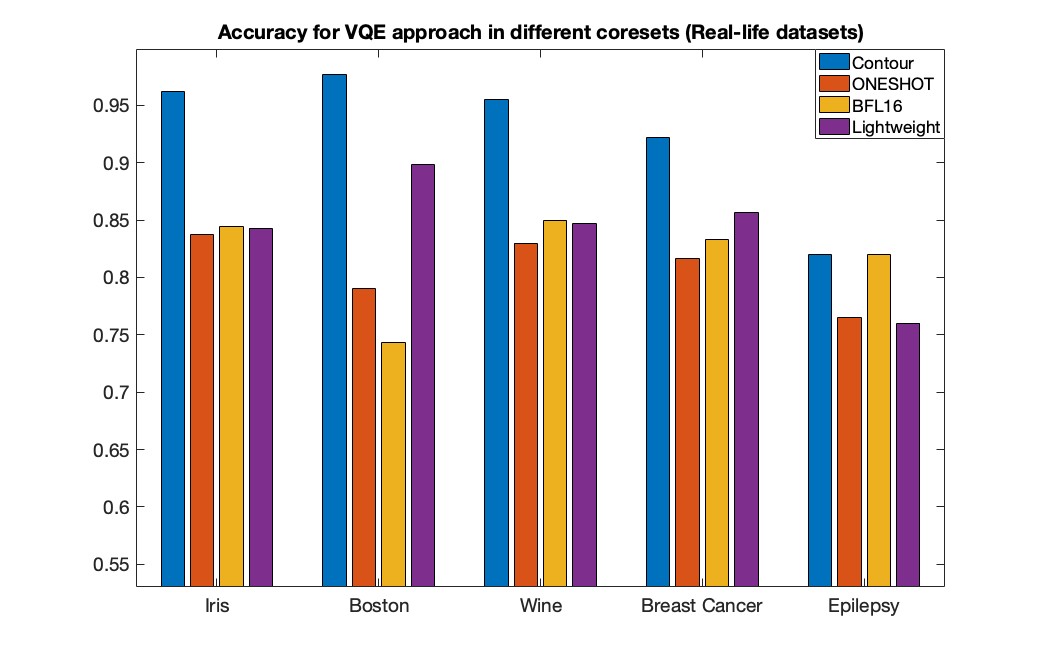}}
    \caption[Accuracy for VQE approach in different coresets (Real-life datasets).]{Accuracy for VQE approach in different coresets (Real-life datasets).}
    \label{coresetRealWorld}
\end{figure}

\begin{figure}[]
    \centering
    \makebox[0pt]{\includegraphics[width=1.2\textwidth]{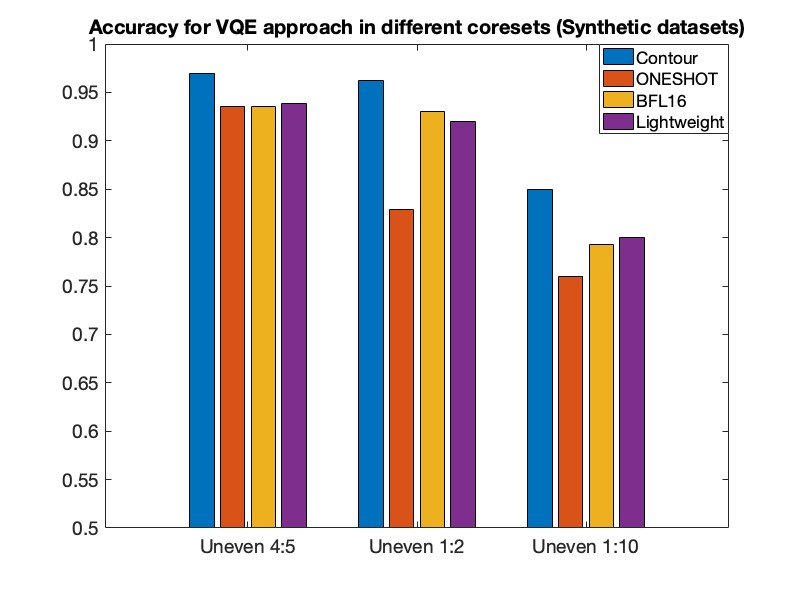}}
    \caption[Accuracy for VQE approach in different coresets (Synthetic datasets).]{Accuracy for VQE approach in different coresets (Synthetic datasets).}
    \label{coresetSyn}
\end{figure}

Our analysis of real-life datasets has shown that the Contour coreset provides the most accurate clustering, surpassing the next best coreset by over 10\% (see Figure \ref{coresetRealWorld}). The minimum accuracies obtained by the Contour coreset in various datasets are also significantly higher than other coresets, particularly in the Iris dataset, with a minimum accuracy of 85.3\% compared to below 60\% for all other coresets. Furthermore, the Contour coreset outperforms other coresets in synthetic datasets with uneven distribution, resulting in significantly higher clustering accuracy ranging from 4\% to over 10\% (Figure \ref{coresetSyn}). Overall, the Contour coreset performs better than other coresets in datasets with uneven data ratios between two clusters.

Moreover, the Contour coreset has a significantly lower standard deviation than other coresets, indicating more consistent and reliable results. The rationale is that the Contour coreset remains the same for a specific dataset, while other coresets generate different coresets each time. 

In our investigation of the Hamiltonian estimated using Taylor approximation, we hypothesized that solving the Hamiltonians formed with higher orders of Taylor Approximation, such as second-order, with VQE would yield more precise clustering results. This would be particularly beneficial for datasets with highly uneven distribution between two clusters. However, our simulations revealed that the VQE circuit measurement resulted in all zeroes when estimating Hamiltonians formed by the second-order Taylor expansion. Other researchers who tried using a brute force approach have also encountered this issue \cite{tomesh2021coreset}. Therefore, we decided to test the Hamiltonian created through the zeroth and first-order Taylor approximation using our VQE circuit.

The VQE circuit's accuracy is significantly improved when employing the Hamiltonian derived from the first-order Taylor approximation compared to the zeroth-order approximation. In particular, the Hamiltonian with first-order Taylor approximation achieves a 5\% higher accuracy when the synthetic data is unevenly distributed. On the other hand, the zeroth-order approximation only reaches about 50\% accuracy for Boston's housing price dataset. In contrast, the first-order approximation achieves nearly 100\%. We conclude that the first-order approximation is more effective for datasets with uneven distribution. Through empirical simulation results, we are the first to prove that solving the first-order Taylor approximated Hamiltonian with VQE can obtain higher accuracy.

\subsection{Optimal VQE setting}

In investigating the most effective configurations for the VQE circuit, we discovered that setting the Contour coreset size to 5 produces stable and accurate clustering results for most datasets. While smaller sizes such as 3 may have comparable performance, the standard deviation of clustering accuracies among datasets is higher. It is also important to note that larger sizes like 10 lead to worse performance due to longer running times and suboptimal labelling. Furthermore, increasing the coreset size beyond 5 can result in decreased accuracy and increased standard deviation of accuracies obtained. This is because VQE struggles to find the ground state of the Hamiltonian as the coreset size increases.

Besides, we found that the Improved Stochastic Ranking Evolution Strategy Optimizer (ISRES) is the optimal optimizer for improving the VQE circuit among all available optimizers provided in Qiskit \cite{Qiskit}. However, note that there is no significant advantage of global optimizers over local optimizers. 

Increasing the number of layers in a VQE circuit does not always result in better accuracy. Larger quantum circuits may introduce more quantum noise, such as gate errors or decoherence. Additionally, running the VQE circuit with more repetitions takes a more extended amount of time. For handling high-dimensional datasets, using fewer layers may be beneficial. This is because increasing the number of layers in VQE only improves accuracy in some cases in our simulations, and increasing the layers is more time-consuming for running the quantum circuit.

In addition, our research has indicated that utilizing the linear entanglement strategy (i.e. qubit \(i\) entangled with qubit \(i+1\) in the entanglement layer) results in the VQE circuit achieving the highest accuracy and producing the lowest standard deviation among real-life datasets when compared to other available entanglement strategies on Quiskit, such as full entanglement, pairwise, and shifted-circular-alternating (SCA) \cite{Qiskit}. However, it is essential to note that the advantage of utilizing linear entanglement is not substantial, with only a 2\% increase in accuracy and a 0.01 decrease in standard deviation.

\begin{figure}[]
    \centering
    \makebox[0pt]{\includegraphics[width=1.2\textwidth]{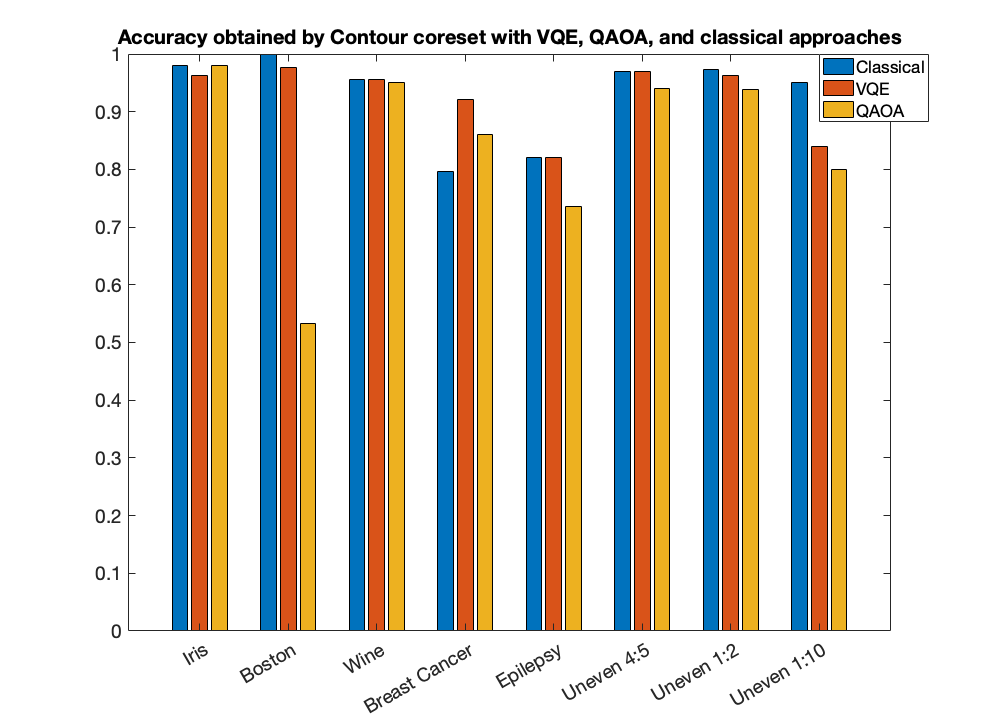}}
    \caption[Accuracy obtained by Contour coreset with VQE, QAOA, and classical approaches]{Accuracy obtained by Contour coreset with VQE, QAOA, and classical approaches}
    \label{VQEQAOAContour}
\end{figure}

Finally, we compared VQE with the optimal settings deduced above, QAOA, and the classical method (applying Lloyd's algorithm on the coreset) implemented by Sklearn (see Figure \ref{VQEQAOAContour}) \cite{scikit-learn}. The VQE approach outperforms the QAOA methods on all datasets except for the Iris dataset, where the difference in accuracy is less than 2\%. One possible explanation for QAOA's better performance on the Iris dataset is that the dataset contains three classes, but we only perform a 2-means clustering. Recall that our clustering accuracy calculation assumes the labels of classical 2-means clustering results as ground truth. However, classical 2-means clustering on the three classes Iris dataset may behave unpredictably since the number of clusters and classes do not match. Therefore, the clustering accuracy calculated for the Iris dataset may not truly reflect the performance of QAOA and VQE. On the other hand, the VQE method achieves almost 100\% accuracy in the Boston housing price dataset. In comparison, the QAOA method only achieves slightly over 50\%. The results show that the VQE method has a significant advantage over QAOA. Furthermore, we observe that the VQE method sometimes outperforms the classical method. The simulations show that when using the Contour coreset, the VQE approach can outperform the classical method in mimicking the result of classical 2-means clustering on the entire dataset.

\subsection{Performance under noisy environment}
Qiskit defines the depolarizing channel as follows \cite{Qiskit}:

$$E(\rho)=(1-\lambda)\rho+\lambda Tr[\rho]\frac{I}{2^{n}}$$

where $\rho$ is the quantum state, $\lambda$ is the depolarizing error parameter with $0 \leq \lambda \leq 4^{n}/(4^{n}-1)$, and $n$ is the number of qubits.

\begin{figure}[]
    \centering
    \makebox[0pt]{\includegraphics[width=1.2\textwidth]{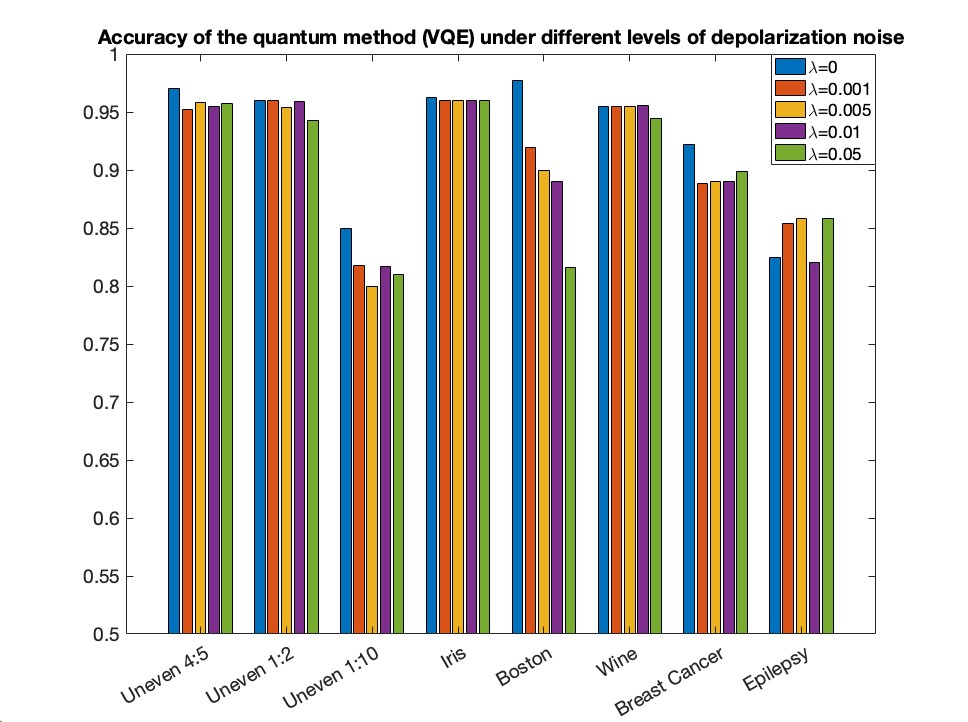}}
    \caption[Accuracy of the quantum method (VQE) under different levels of depolarization noise]{Accuracy of the quantum method (VQE) under different levels of depolarization noise.}
    \label{deplonoiseAcc}
\end{figure}

We analyzed the effectiveness of the VQE and Contour coreset approach in dealing with depolarization noise at various levels (Figure \ref{deplonoiseAcc}). Accuracy remained consistent within 5\% differences under depolarisation noise across most datasets, with only two exceptions: the Boston housing price and Epilepsy datasets. In the case of the Boston dataset, the clustering accuracy was reduced by more than 15\% under depolarization noise. We discovered this was due to a non-optimal partition state for the coreset points, leading to an energy level close to the ground level of the Hamiltonian with a clustering accuracy of only 53\%. This alternative partition dragged down the overall accuracy as the level of depolarization noise increased. However, for the Epilepsy dataset, the clustering accuracy surprisingly increased with the level of depolarization noise. We believe this is because the Contour coreset could not accurately approximate the entire dataset with numerous instances and attributes (11500 instances and 179 attributes). Therefore, the ground level converged by VQE did not match the optimal clustering partition. Noteworthy, the standard deviation of all datasets remained consistent even with increased noise levels. Hence, we conclude that combining VQE and the Contour coreset can effectively handle depolarization noise with minimal impact on clustering performance.

\begin{figure}[]
    \centering
    \makebox[0pt]{\includegraphics[width=1.2\textwidth]{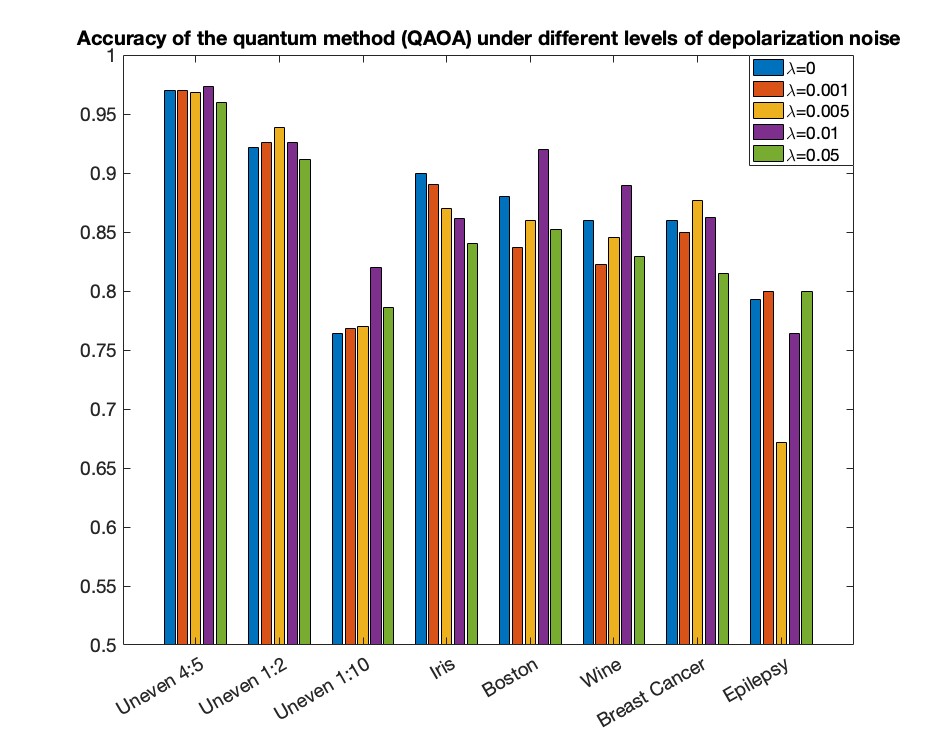}}
    \caption[Accuracy of the quantum method (QAOA) under different levels of depolarization noise]{Accuracy of the quantum method (QAOA) under different levels of depolarization noise.}
    \label{deplonoiseAccQAOA}
\end{figure}

On the other hand, results show that the QAOA method with Contour coreset has varying accuracies when simulated to depolarization noise (Figure \ref{deplonoiseAccQAOA}). As the noise level increases, the QAOA's accuracy fluctuates, with differences ranging from 5\% to 10\% consistently for all datasets. In general, VQE achieves higher clustering accuracy than QAOA across all datasets and levels of depolarization noise. Moreover, it is surprising that the accuracy obtained by QAOA sometimes increases with increasing noise levels. In most datasets, the noiseless environment does not obtain the highest accuracy. This observation suggests that the partition process in QAOA contains randomness, making optimizing and explaining the method challenging. Furthermore, the standard deviation of QAOA's accuracy is higher than that of VQE, with a maximum difference of 0.1 between noise levels. We conclude that VQE is more reliable in the presence of depolarization noise than QAOA.

\section{Discussion and outlook}\label{discussion}

Our study confirms that the proposed approach based on the Contour coreset and VQE outperforms QAOA and other coresets in clustering accuracy and standard deviation in real-life and synthetic datasets, even in highly uneven data distributions. VQE circuit can also utilize the first-order Taylor expansion Hamiltonian for improved accuracy, while QAOA is limited to zeroth-order Taylor approximation due to current quantum hardware limitations, such as the challenge of supporting a high degree of connectivity among qubits \cite{tomesh2021coreset}. Moreover, the Contour coreset has a shorter construction time than other methods, even for large and high-dimensional datasets. Lastly, VQE maintains higher clustering accuracy and lower standard deviation in the presence of depolarization noise compared to QAOA.

When dealing with large datasets, the VQE can be improved as the standard deviation tends to increase to approximately 0.1. Besides, the performance of VQE fluctuates under varying levels of depolarization noise in some datasets. Thus, further research is required to improve the handling of datasets with high volumes of data and high dimensions.

We have developed the Contour coreset algorithm by addressing existing coreset algorithms' limitations and demonstrated our coreset's efficiency based on simulation results. However, we did not provide rigorous mathematical proof for the performance bound of the Contour coreset, e.g., the lower bound of the worst-case performance. Previous research on coreset construction algorithms typically includes an accuracy boundary that correlates with the size of the coreset or the number of clusters for partition \cite{bachem2017practical, bachem2018one, bachem2018scalable}. Establishing mathematical proof that supports and explains the behaviour of the Contour coreset will open the opportunity to improve the Contour coreset with a theoretically based direction rather than using a guesswork approach.

\section{Conclusion}\label{conclusion}

We developed the Contour coreset specifically for solving the 2-means clustering problem with variational quantum algorithms. It outperformed existing coresets by taking less time to construct and can approximate datasets with limited points. We successfully identified the optimal settings for the VQE circuit in solving the clustering problem, including developing Hamiltonian with the first-order Taylor approximation. Our proposed quantum approach that combines the Contour coreset and VQE achieves higher accuracy and lower standard deviation than existing quantum approaches with QAOA and standard coresets. 

\appendix

\section{Background definitions}

\subsection{Coreset}

Coreset is a data reduction technique originated from geometric approximation in computational geometry \cite{agarwal2005geometric}. It is a smaller set of weighted data points representing the original dataset, which can be utilized for computations and data analysis  \cite{harrow2020small, tomesh2021coreset, qu2022performance}. Advanced coreset construction methods like BFL16, BFL17, and ONESHOT use concepts such as importance sampling, sensitivity, and $D^2$-sampling \cite{bachem2017practical, bachem2018one}. 

\subsubsection{\texorpdfstring{($\epsilon,k$)-coreset}\ }

($\epsilon,k$)-coreset is mathematical property held by some coresets developed for k-means clustering. It provides a theoretical guarantee for the coreset's performance on approximating the full dataset \cite{bachem2018one, lucic2016strong, bachem2018scalable}. The mathematical definition is as follows: 

\begin{definition}[($\epsilon,k$)-coreset]
A weight set $C$ is a ($\epsilon,k$)-coreset for X if $\forall Q \subset \mathbb{R}, |Q|\leq k$
$$|\phi_{X}(Q)-\phi_{C}(Q)|\leq \epsilon \phi_{X}(Q)$$
where $\phi_{X}$ is the quantization error for k-means clustering as described above.
\end{definition}
The $\phi_{C}(Q)$ can approximate $\phi_{X}(Q)$ up to $1 \pm \epsilon$ multiplicative factor for all possible sets of cluster centers simutaneously \cite{bachem2018scalable}. For example, if C is a ($\epsilon,k$)-coreset of $X$ with $\epsilon \in (0, 1/4)$, we have $\phi_{X}(Q_{C}^{*}) \leq (1+4\epsilon)\phi_{X}(Q_{X}^{*})$, where $Q^{*}$ denotes the optimal solution with k centers on the respective set of data.

\subsubsection{Importance sampling, Sensitivity, and $D^{2}$ sampling}

The importance sampling technique is an entrenched approach for approximately sampling important points with the unbiased estimation of cost function \cite{bachem2017practical}. The linearity of expectation can prove the unbiasedness of importance sampling. The cost function here refers to the objective function the machine learning problem aims to minimize.

Researchers have studied the role of sensitivity in importance sampling to minimize the estimator's variance through a sampling distribution. Sensitivity refers to the impact of data points on the objective function in the worst-case scenario \cite{langberg2010universal}. In order to decrease the variance of the estimator through a sampling distribution, researchers proposed the $D^2$-sampling method \cite{langberg2010universal, arthur2006k}. This method prioritizes time efficiency over solution quality when approximating the optimal solution, which serves as an estimate for the worst-case impact bound.

\subsubsection{Coreset construction: Lightweight, BFL16 and ONESHOT}

Lightweight coreset is a compact coreset method developed for k-means clustering that uses a parallel algorithm to summarize the dataset in just two passes \cite{bachem2018scalable}. It can handle additive and multiplicative approximation errors based on importance sampling. An advantage of this method is that all points will be sampled with nonzero probability as candidates of coresets. Moreover, it prioritizes those that differ from the dataset's average as coreset points, which enhances the precision of clustering.

The BFL16 algorithm is primarily designed for k-means clustering. It can also be applied to other machine learning tasks, such as k-median clustering and Principal Component Analysis (PCA). This algorithm builds a coreset by using k-means ++ and $D^{2}$ sampling techniques. Previous studies have combined BFL16 with QAOA to address the 2-means clustering problem \cite{qu2022performance, tomesh2021coreset}.

ONESHOT coreset is a versatile coreset for k-clustering problems that utilizes k-means ++ and $D^{2}$ sampling techniques, similar to BFL16 \cite{bachem2018one}. Moreover, the coreset's gradient sensitivity is regulated by a hyper-parameter $\Delta$, in which $\Delta$ bounds the sensitivity with the exponential grid. It works well in both metric and Euclidean spaces. Research has indicated that when combined with QAOA, ONESHOT produces more favorable clustering outcomes than other coreset options \cite{qu2022performance}.

\subsection{2-means clustering with MAX-CUT and QAOA}

The core idea is mathematically transforming the 2-means clustering problem into a weighted MAX-CUT problem and solving it with QAOA \cite{tomesh2021coreset, harrow2020small, qu2022performance}. The procedures are as follows:

\begin{enumerate}
   \item[Step 1:] Construct coresets from the large classical dataset. Past literature has used coresets developed for k-means clustering, such as BFL16 and ONESHOT \cite{tomesh2021coreset, harrow2004superdense}.
   \item[Step 2:] Transform the objective function of the 2-means clustering problem into a weighted MAX-CUT problem. The mathematical proof is included in \ref{maths proof for 2-means to max cut}. 
   \item[Step 3:] Obtain the Hamiltonian of the MAX-CUT problem by Taylor Approximation. Past literature has shown experimental results in the zeroth-order Taylor Approximation, which can optimally approximate a balanced dataset, i.e., the number of data in two labels equals each other \cite{tomesh2021coreset}. The zeroth-order Taylor Approximation can reasonably estimate datasets with unequal amounts of data in two labels if the ratio does not exceed 3:2. The mathematical proof is included in \ref{maths proof for 2-means to max cut}.
   \item[Step 4:] Find the ground state with QAOA. Researchers also examine the performance of quantum autoencoders in restoring the error states to pure states due to the depolarisation quantum noise \cite{qu2022performance}.
   \item[Step 5:] Data labels assignment. The result obtained from QAOA shows the partition that maximizes inter-cluster distance. We will first calculate the centroids of the two clusters based on the partitioned coreset points. Then we can assign each classical data to its closest centroid and assign the label.
\end{enumerate}

This approach has achieved around 80-90\% accuracy in real-life data \cite{harrow2020small, tomesh2021coreset, qu2022performance}. However, it only applies to 2-means clustering due to the MAX-CUT problem. Corset constructions using advanced methods such as importance sampling and sensitivity bounding do not result in significant accuracy improvement, and there is no clear relationship between coreset size and accuracy achieved. While QAOA has shown promise, excessive iterations are often necessary to achieve around 90\% accuracy, and there is still room for improvement in obtaining the best partition.

\section{Maths proof for converting 2-means clustering to MAX-CUT problem and the corresponding Hamiltonian}

\label{maths proof for 2-means to max cut}
Note that the mathematical proof and algebraic annotations are referenced from \cite{tomesh2021coreset}. The derivation up to the first-order Taylor approximation is taken from \cite{tomesh2021coreset} and the second-order Taylor approximation is originally derived.

To begin with, we denote the coreset size as $m$, data points in dataset as $\mathbf{x}_{1}, \dots, \mathbf{x}_{n} \in \mathbb{R}$, the two cluster centers as $\mathbf{\mu}_{-1}$ and $\mathbf{\mu}_{+1}$, and the two clustering sets as $S_{-1}$ and $S_{+1}$. Now, we consider the objective function of 2 means clustering:

\begin{equation}
\label{eqn3}
\sum_{i\in S_{-1}}w_{i}\|\mathbf{x}_{i}-\mathbf{\mu}_{-1}\|^{2}+\sum_{i\in S_{+1}}w_{i}\|\mathbf{x}_{i}-\mathbf{\mu}_{+1}\|^{2}
\end{equation}

where the cluster centers are:

$$\mathbf{\mu}_{-1} = \frac{\sum_{i\in S_{-1}}w_{i}\mathbf{x}_{i}}{W_{-1}} \text{\, and \,} \mathbf{\mu}_{+1} = \frac{\sum_{i\in S_{+1}}w_{i}\mathbf{x}_{i}}{W_{+1}}$$

By the Law of Total Variance and mathematically proofs in \cite{kriegel2017black} and \cite{bauckhage2017adiabatic}, minimizing \ref{eqn3} equals to maximizing the weighted intercluster distance

\begin{equation}
\label{eqn4}
W_{+1}W_{-1}\|\mathbf{\mu}_{+1}-\mathbf{\mu}_{-1}\|^{2}
\end{equation}

We also define 

$$W_{\pm1} = \sum_{i\in S_{\pm1}}w_{i} \text{\, and \,} W \equiv W_{-1} + W_{+1}$$

To begin with, we consider the two optimal clusters holds equal weights. Therefore we have $W_{+1} = W_{-1}$ and $W_{+1}W_{-1}$ are maximized by $W_{+1} = W_{-1} = W/2$

Thus, we can reconsider \ref{eqn4} as:

\begin{flalign*}
& \left( \frac{W}{2}\right)^2\bigg\| \frac{\sum_{i\in S_{-1}}w_{i}\mathbf{x}_{i}}{W/2} - \frac{\sum_{i\in S_{+1}}w_{i}\mathbf{x}_{i}}{W/2}\bigg\|^2 = \bigg\| \sum_{i\in S_{-1}}w_{i}\mathbf{x}_{i} - \sum_{i\in S_{+1}}w_{i}\mathbf{x}_{i}\bigg\|^2&\\
     & = \sum_{i}w_{i}^2\|\mathbf{x}_{i}\|^2 + 2\sum_{i<j\in S_{-1}}w_{i}w_{j}\mathbf{x}_{i}\cdot\mathbf{x}_j +2\sum_{i<j\in S_{+1}}w_{i}w_{j}\mathbf{x}_{i}\cdot\mathbf{x}_j - 2\sum_{i\in S_{-1},j\in S_{+1}}w_{i}w_{j}\mathbf{x}_{i}\cdot\mathbf{x}_j&\\
     & = \sum_{i}w_{i}^2\|\mathbf{x}_{i}\|^2 + 2\sum_{i<j}w_{i}w_{j}\mathbf{x}_{i}\cdot\mathbf{x}_j - 4\sum_{i\in S_{-1},j\in S_{+1}}w_{i}w_{j}\mathbf{x}_{i}\cdot\mathbf{x}_j
\end{flalign*}

Note that only the term \ $4\sum_{i\in S_{-1},j\in S_{+1}}w_{i}w_{j}\mathbf{x}_{i}\cdot\mathbf{x}_j$ \ depends on the partitioning by $S_{-1} \text{\,and\,} S_{+1}$, so we can reformulate the 2-means clustering objective to 

\begin{equation}
\label{eqn5}
\sum_{i\in S_{-1}, j \in S_{+1}} - w_{i}w_{j}\mathbf{x}_{i}\cdot\mathbf{x}_j    
\end{equation}

Now, in order to maximize \ref{eqn5} with QAOA or VQE, we have to transform it as a Hamiltonian. Consider the Pauli Z operators as $Z_{i}, Z_{j} \in \{-1, +1\}$, where $\frac{1-Z_{i}Z_{j}}{2} = 1$ for $Z_{i} \neq Z_{j}$ and equals to 0 for $Z_{i} = Z_{j}$. \ref{eqn5} can be represented as the following Hamiltonian:

\begin{equation}
\label{eqn6}
H_{p} = \sum_{i<j}w_{i}w_{j}\mathbf{x}_{i}\cdot\mathbf{x}_j Z_{i}Z_{j}
\end{equation}

However, most of the time we do not get the optimal case scenario where the clusters' weights equal to each other. Therefore, for unequal cluster weights, we reconsider \ref{eqn4} as:

\begin{equation}
\begin{aligned}[b]
&W_{-1}W_{+1}\|\mathbf{\mu}_{-1}-\mathbf{\mu}_{+1}\|^{2} &\\
&= \bigg\| \frac{\sqrt{W_{-1}W_{+1}}}{W_{-1}}\sum_{i\in S_{-1}} w_{i}\mathbf{x}_{i} - \frac{\sqrt{W_{-1}W_{+1}}}{W_{+1}}\sum_{i\in S_{+1}} w_{i}\mathbf{x}_{i} \bigg\|^2 &\\
& = \bigg\| \frac{\sqrt{W_{+1}}}{\sqrt{W_{-1}}}\sum_{i\in S_{-1} w_{i}\mathbf{x}_{i}} -  \frac{\sqrt{W_{-1}}}{\sqrt{W_{+1}}}\sum_{i\in S_{+1} w_{i}\mathbf{x}_{i}} \bigg\|^2 &\\
& = \sum_{i}\biggl( \frac{W_{+1}}{W_{-1}} \text{if $i \in S_{-1}$, else} \frac{W_{-1}}{W_{+1}}\biggr)w_{i}^2\|\mathbf{x}_{i}\|^2 &\\
&\quad  +2\sum_{i<j}( \frac{W_{+1}}{W_{-1}} \text{if $i,j \in S_{-1}$,} \frac{W_{-1}}{W_{+1}} \text{if $i,j \in S_{+1}$, else -1}\biggr) w_{i}w_{j}\mathbf{x}_{i}\cdot\mathbf{x}_j
\end{aligned}
\label{eqn7}
\end{equation}

We now examine these two terms in \ref{eqn7} that represent the ratio between the two clusters' weights:

\begin{equation}
    \frac{W_{+1}}{W_{-1}} = \frac{1}{W_{-1}/W}-1 \text{\, and\, } \frac{W_{-1}}{W_{+1}} = \frac{1}{W_{+1}/W}-1
\label{eqn8}
\end{equation}

We will perform the Taylor expansion of these ratios around $x\equiv W_{-1}/W = W_{+1}/W = 1/2$, which follows the equal cluster weight scenario. The motivation behind this approximation is that we can formulate the resulting polynomial as Hamiltonian and presented as $Z_{i}$ terms. 

The zeroth order Taylor approximation gives $1/x\approx 2$ and the first order gives $4-4x$. The Hamiltonian formed by the zeroth order Taylor expansion is the same as the one presented in \ref{eqn6}.

For the first order Taylor expansion, we expand \ref{eqn8} and get:

$$\frac{W_{+1}}{W_{-1}} \approx 3-\frac{4}{W}W_{-1} \text{\, and \,} \frac{W_{-1}}{W_{+1}} \approx 3-\frac{4}{W}W_{+1}$$

By transforming into Pauli Z operators, we obtain:

$$\frac{W_{+1}}{W_{-1}} \approx 1+\frac{2}{W}\sum_{l}w_{i}Z_{l} \text{\, and \,} \frac{W_{-1}}{W_{+1}} \approx 1-\frac{2}{W}\sum_{l}w_{i}Z_{l}$$

With these ratio terms, we reconsider \ref{eqn7}:

\begin{flalign*}
&W_{-1}W_{+1}\|\mathbf{\mu}_{-1}-\mathbf{\mu}_{+1}\|^{2} &\\
& = \sum_{i} \biggl(1+\frac{2}{W}\sum_{l}w_{l}Z_{l} \text{\, if \, } i\in S_{-1} \text{\, ,else\, } 1-\frac{2}{W}\sum_{l}w_{l}Z_{l} \biggr)w_{i}^2\| \mathbf{x}_{i}\|^2 &\\
& \quad  + 2\sum_{i<j} \biggl(1+\frac{2}{W}\sum_{l}w_{l}Z_{l} \text{\, if \, } i, j \, \in S_{-1} \text{\, ,} 1-\frac{2}{W}\sum_{l}w_{l}Z_{l} \text{\,if} i, j \, \in S_{+1} \text{, else -1} \biggr)w_{i}w_{j}\mathbf{x}_{i} \cdot \mathbf{x}_{j}
\end{flalign*} 

The indicator functions can be further formulated as Pauli Z operators as follows:

\begin{equation}
\begin{aligned}[b]
& \sum_{i} \biggl( 1-\frac{2Z_{i}}{W}\sum_{l}w_{l}Z_{l} \biggr) w_{i}^2\| \mathbf{x}_{i} \|^2 + 2\sum_{i<j} \biggl( Z_{i}Z_{j} - \frac{Z_{i}+Z_{j}}{W}\sum_{l} w_{l}Z_{l}\biggr)w_{i}w_{j}\mathbf{x}_{i} \cdot \mathbf{x}_{j} &\\
& = \sum_{i} \biggl( 1-\frac{2Z_{i}}{W}\sum_{j \neq i}^{m} w_{j} Z_{j} - \frac{2w_{i}}{W}\biggr)w_{i}^{2} \| \mathbf{x}_{i}\|^2 &\\
& \quad + 2\sum_{i<j}\biggl( Z_{i}Z_{j} - \frac{1}{W}(Z_{i}\sum_{k \neq i}^{m} w_{k}Z_{k} + Z_{j}\sum_{k \neq j}^{m} w_{k}Z^{k} + w_{i} + w_{j}) \biggr) w_{i}w_{j}\mathbf{x}_{i} \cdot \mathbf{x}_{j}
\label{eqn9}
\end{aligned}
\end{equation}

Note that the $m$ in \ref{eqn9} is the total number of qubits using.

We have further calculated the second order Taylor Expansion, where $x\approx8x^2-12x+6$

Consider $\frac{W_{+1}}{W_{-1}} = \frac{1}{W_{-1}/W}-1 \, , \, \frac{W_{-1}}{W_{+1}} = \frac{1}{W_{+1}/W}-1$

By substitution of the second order Taylor expansion, we have:

$$\frac{W_{+1}}{W_{-1}} \approx 8 \biggl( \frac{W_{-1}}{W}\biggr)^2 - 12 \biggl(\frac{W_{-1}}{W}\biggr) +6 -1 = 8 \biggl( \frac{W_{-1}}{W}\biggr)^2 - 12 \biggl(\frac{W_{-1}}{W}\biggr) +5 $$
$$\frac{W_{-1}}{W_{+1}} \approx 8 \biggl( \frac{W_{+1}}{W}\biggr)^2 - 12 \biggl(\frac{W_{+1}}{W}\biggr) +5 $$

In order to express the terms in Pauli Z operators for the Hamiltonian form, first we express $W_{-1}$ and $W_{+1}$ as follows:

$$W_{-1} = \frac{W-\sum_{l}w_{l}Z_{l}}{2} \, , \, W_{+1} = \frac{W+\sum_{l}w_{l}Z_{l}}{2}$$

Thus, by substitution, we have:

\begin{equation}
\begin{aligned}[b]
& \frac{W_{+1}}{W_{-1}} &\\
& = 8 \biggl( \frac{1}{W} (\frac{W-\sum_{l}w_{l}Z_{l}}{2})\biggr)^2 - 12\biggl( \frac{1}{W} (\frac{W-\sum_{l}w_{l}Z_{l}}{2})\biggr) +5 &\\
&= \frac{2}{W^2}(W-\sum_{l}w_{l}Z_{l})^2 - \frac{6}{W}(W-\sum_{l}w_{l}Z_{l}) + 5 &\\
& = \frac{2}{W^2}\biggl( W^2-2W\sum_{l}w_{l}Z_{l} + (\sum_{l}w_{l}Z_{l})^2\biggr)-6+\frac{6}{W}\sum_{l}w_{l}Z_{l}+5 &\\
& = 2-\frac{4}{W}\sum_{l}w_{l}Z_{l} + \frac{2}{W^2}(\sum_{l}w_{l}Z_{l})^2 + \frac{6}{W}\sum_{l}w_{l}Z_{l}-1 &\\
& = \frac{2}{W^2}(\sum_{l}w_{l}Z_{l})^2 + \frac{2}{W}\sum_{l}w_{l}Z_{l} +1
\label{eqn10}
\end{aligned}
\end{equation}

Similarly, 

\begin{equation}
\begin{aligned}[b]
& \frac{W_{-1}}{W_{+1}} &\\
&= 8 \biggl( \frac{1}{W} (\frac{W+\sum_{l}w_{l}Z_{l}}{2})\biggr)^2 - 12\biggl( \frac{1}{W} (\frac{W+\sum_{l}w_{l}Z_{l}}{2})\biggr) +5 &\\
&= \frac{2}{W^2}(W+\sum_{l}w_{l}Z_{l})^2 - \frac{6}{W}(W+\sum_{l}w_{l}Z_{l}) + 5 &\\
& = \frac{2}{W^2}\biggl( W^2+2W\sum_{l}w_{l}Z_{l} + (\sum_{l}w_{l}Z_{l})^2\biggr)-6-\frac{6}{W}\sum_{l}w_{l}Z_{l}+5 &\\
& = 2+\frac{4}{W}\sum_{l}w_{l}Z_{l} + \frac{2}{W^2}(\sum_{l}w_{l}Z_{l})^2 - \frac{6}{W}\sum_{l}w_{l}Z_{l}-1 &\\
& = \frac{2}{W^2}(\sum_{l}w_{l}Z_{l})^2 - \frac{2}{W}\sum_{l}w_{l}Z_{l} +1
\label{eqn11}
\end{aligned}
\end{equation}

Thus, when we reconsider \ref{eqn7}, we have:

\begin{equation}
\begin{aligned}[b]
& W_{-1}W_{+1}\|\mathbf{\mu}_{-1}-\mathbf{\mu}_{+1}\|^{2} &\\
& = \sum_{i} \biggl( \ref{eqn10} \text{\, if \,} i\in S_{-1}, \text{else} \ref{eqn11} \biggr)w_{i}^2\|\mathbf{x}_{i}\|^2 &\\
& \quad + 2 \sum_{i<j} \biggl( \ref{eqn10} \text{\, if \,} i,j \in S_{-1}, \ref{eqn11} \text{if \,}  i,j \in S_{+1} \text{\, , else -1}\biggr)w_{i}w_{j}\mathbf{x}_{i} \cdot \mathbf{x}_{j} &\\
& = \sum_{i} \biggl( \frac{2}{W^2}(\sum_{i} w_{l}Z_{l})^2 - \frac{2Z_{i}}{W} \sum_{i} w_{l}Z_{l} +1\biggr)w_{i}^2\|\mathbf{x}_{i}\|^2 &\\
& \quad + 2\sum_{i<j}\biggl( Z_{i}Z_{j} + \frac{(1+Z_{i}Z_{j})}{W^2}(\sum_{k}w_{k}^2 + 2\sum_{k<l} w_{k}w_{l}Z_{k}Z_{l}) - \frac{Z_{i}+Z_{j}}{2}\sum_{l}w_{l}Z_{l}\biggr)w_{i}w_{j}\mathbf{x}_{i} \cdot \mathbf{x}_{j} &\\
& = \sum_{i} \biggl( \frac{2}{W^2}(\sum_{l}w_{l}Z_{l})^2  - \frac{2Z_{i}}{W}\sum_{l}w_{l}Z_{l} +1\biggr)w_{i}^2\|\mathbf{x}_{i}\|^2 &\\
& \quad +2\sum_{i<j} \biggl( Z_{i}Z_{j} + \frac{1+Z_{i}Z_{j}}{W^2} (\sum_{k}w_{k}^2 + 2\sum_{k<l} w_{k}w_{l} Z_{k}Z_{l}) - \frac{Z_{i}+Z_{j}}{W}\sum_{l}w_{l}Z_{l}  \biggr)w_{i}w_{j}\mathbf{x}_{i} \cdot \mathbf{x}_{j}
\label{eqn12}
\end{aligned}
\end{equation}


\printbibliography
\end{document}